# Strain-driven switching between antiferromagnetic states in frustrated antiferromagnet UO$_2$ probed by exchange bias effect

E. A. Tereshina-Chitrova[1*], L. V. Pourovskii[2,3], S. Khmelevskyi[4], L. Horak[5], Z. Bao[6], A. Mackova[7,8], P. Malinsky[7,8], T. Gouder[9], R. Caciuffo[9,10]

[1]*Institute of Physics, Czech Academy of Sciences, 18121 Prague, Czech Republic, e-mail: teresh@fzu.cz*

[2]*CPHT, CNRS, École polytechnique, Institut Polytechnique de Paris, 91120 Palaiseau, France*

[3]*Collège de France, Université PSL, 11 place Marcelin Berthelot, 75005 Paris, France*

[4]*Vienna Scientific Cluster Research Center, Vienna Technical University, Wiedner Hauptstrasse 8-10, 1040, Vienna, Austria*

[5]*Faculty of Mathematics and Physics, Charles University, 12116 Prague, Czech Republic*

[6]*Malvern Panalytical B.V., Lelyweg 1 7602 EA, Almelo, Overijssel Netherlands*

[7]*Nuclear Physics Institute CAS, Department of Neutron Physics, 25068 Řež, Czech Republic*

[8]*Department of Physics, Faculty of Science, J. E. Purkinje University, 40096 Usti nad Labem, Czech Republic*

[9]*European Commission, Joint Research Centre (JRC), Postfach 2340, DE-76125 Karlsruhe, Germany*

[10]*National Institute for Nuclear Physics, Genoa Section, Genova, I-16146, Italy*

**Abstract:**

Frustrated antiferromagnets offer a captivating platform to study the intricate relationship of magnetic interactions, geometric constraints, and emergent phenomena. By controlling spin orientations, these materials can be tailored for applications in spintronics and quantum information processing. The research focuses on the interplay of magnetic and exchange anisotropy effects in artificial heterostructures based on a canonical frustrated antiferromagnet, UO$_2$. The potential to manipulate the spin directions in this material and switch between distinct antiferromagnetic states is investigated using substrate-induced strain. The phenomenon is probed using exchange bias (EB) effects in stoichiometric UO$_2$/Fe$_3$O$_4$ bilayers. By employing many-body first-principles calculations magnetic configurations in the UO$_2$ layers are identified. Even a minor tetragonal distortion triggers a transition between antiferromagnetic states of different symmetries, driven by a robust alteration of single-ion anisotropy due to the distortion. Consequently, this change influences the arrangement of magnetic moments at the UO$_2$/Fe$_3$O$_4$ interface, affecting the magnitude of exchange bias. The findings showcase how epitaxial strain can effectively manipulate the antiferromagnetic states in frustrated antiferromagnets by controlling single-site anisotropy.

## 1. Introduction

In materials with magnetic atoms sitting in geometrically frustrated lattices, multiple ordered states exhibiting distinct symmetries can possess comparable energies related to inter-site antiferromagnetic (AFM) exchange interactions. The concept of geometric frustration frequently intersects with broader notions in condensed matter physics encompassing quantum spin liquids, topological states, and critical behaviour [1-3]. Over recent decades, the deliberate manipulation of magnetic states in collinear antiferromagnets has garnered considerable attention, primarily driven by advancements in AFM spintronics [4,5]. Hence, research efforts have been directed towards the controlled manipulation of antiferromagnets using diverse methods, including electrical means [6-8], structural alterations [9-11], high magnetic fields [12], and interaction with neighbouring ferromagnetic layers [13].

Frustrated antiferromagnets, while inherently fascinating [14], have yet to gain practical significance. Exploring ways to control single-site anisotropy in these materials, possibly through strain engineering, could boost their practical applications. To illustrate the principle of switching between antiferromagnetic states of *different* symmetries in frustrated magnets, we utilize uranium dioxide (UO$_2$), a principal nuclear fuel material, as a model system. In UO$_2$'s face-centered cubic (*fcc*) lattice, magnetic frustration arises from the energy degeneracy of three antiferromagnetic states (Fig. 1*a*(bottom)), influenced by relativistic inter-site AFM exchange interactions [15]. The nature of UO$_2$'s magnetic ground state has puzzled scientists for decades [16]. Initial neutron diffraction measurements in the 1960s suggested a collinear 1***k*** antiferromagnetic [17,18] structure (Fig. 1*a*(top)) below the first-order transition at the Néel temperature $T_N$ = 30.8 K. In the 1970s, the occurrence of a dynamical Jahn-Teller (J-T) distortion above the phase transition temperature was proposed [19]. Subsequent neutron diffraction studies unveiled the condensation of a static J-T distortion at T$_N$, hinting at a noncollinear 2***k*** AFM order in UO$_2$ [20,21]. Eventually [22], the exotic transverse 3***k*** magnetic ordering, a superposition of three individual 1k structures in the cubic unit cell, with uranium magnetic moments of approximately 1.74(2) μ$_B$ pointing in the <111> directions (Fig. 1*a*(bottom)) was revealed. Resonant x-ray scattering experiments further provided evidence [23] of long-range antiferro-ordering of electric-quadrupole moments at the uranium site.

The ground state triplet of the $U^{4+}$ ion in the cubic crystal field of $UO_2$ is spherically symmetric, resulting in exactly zero single-ion anisotropy in the cubic phase [15]. The energy degeneracy between the states within the *fcc* lattice is lifted [24,25] by higher rank inter-site exchange interactions of relativistic origin that couple quadrupole moments and stabilize [24] the 3*k* order in cubic $UO_2$. Here, we demonstrate that an effective switching between the AFM states of different symmetries can be achieved by stretching the lattice of $UO_2$ and inducing strong single-ion anisotropy.

In $UO_2$ thin films strain can induce ferromagnetism by incorporating point defects and generating hypo- ($x < 0$) or hyper- ($x > 0$) stoichiometric $UO_{2+x}$ [26]. In this study, we apply epitaxial strain through substrate selection and switch between antiferromagnetic structures in stoichiometric $UO_2$ films. Identifying the exact type of antiferromagnetic ordering in $UO_2$ films is challenging. Conventional laboratory macroscopic magnetometry probes are unsuitable due to insensitivity of antiferromagnets to external magnetic fields, whereas neutron diffraction requires larger volume of material. Instead, we utilize the exchange-bias (EB) [27] effect's amplitude in $UO_2$/$Fe_3O_4$ bilayers to indirectly probe the antiferromagnetic state in $UO_2$. The EB amplitude significantly varies depending on the substrate used for growing the $UO_2$/$Fe_3O_4$ bilayers. Employing many-body first-principles calculations based on charge self-consistent DFT+Hubbard-I (HI) approximation [28-30] and a force-theorem-HI (FT-HI) approach [31] for inter-site exchange, we predict that tetragonal distortion leads to a transition in $UO_2$ from transverse 3*k* magnetic order to transverse 1*k* magnetic order. The experimental observation of exchange bias in the bilayers $UO_2$/$Fe_3O_4$ supports the notion that epitaxial strain serves as a crucial control parameter in determining the $UO_2$'s antiferromagnetic structure.

## 2. Results and discussion

*1. Structural and magnetic characterization of thin films*

We have previously observed [32] a significant exchange bias effect (EB) in LAO-based $UO_2$/$Fe_3O_4$ bilayers when they were field-cooled below the Néel temperature of $UO_2$. The EB arises from the interfacial exchange coupling between the antiferromagnetic $UO_2$ and the ferrimagnetic $Fe_3O_4$, leading to a shift in the magnetic hysteresis loop. While exchange bias is typically attributed to interface phenomena, previous studies have indicated [33] that the mechanisms behind EB, which are not yet fully understood, can also be influenced by the spin configurations of the bulk antiferromagnetic material. In our current study, we utilized this property to investigate the dependence of EB on the thickness of $UO_2$ while introducing the strain in $UO_2$.

For the samples preparation we have chosen the substrates due to the following considerations. $UO_2$ has a lattice constant $a_0$ of 5.469 Å at room temperature. $CaF_2$ is isostructural to $UO_2$ with lattice parameter $a_0 = 5.462$ Å (lattice mismatch of 0.1%). For the (001) $LaAlO_3$ substrates with lattice parameter $a_0 = 3.821$ Å, the epitaxial relationship with $UO_2$ would be such that the (110) plane of $UO_2$ (*d*-spacing $a_0/\sqrt{2} = 3.867$ Å) fits with the $LaAlO_3$ (100) plane, thus producing a small compression of $UO_2$ of −1.2% with respect to the substrate in-plane spacing. Indeed, a XRD study confirmed that the $UO_2$ layers grow in the [100] direction on both types of substrates, $CaF_2$ (100) and $LaAlO_3$ (001). Bulk magnetite $Fe_3O_4$ (*fcc* structure with space group Fd3m) has a lattice parameter of 8.39 Å. We find that $Fe_3O_4$ grows on $UO_2$ with an [111]-out-of plane orientation (Fig. 1(*b*)). Figures 1 (*c*) and (*d*) demonstrate the stacking sequence for the $UO_2$/$Fe_3O_4$ bilayers on substrates (100) $CaF_2$ and (001) $LaAlO_3$, respectively (the hexagonal Mg capping layer with basal planes parallel to the surface is not shown).

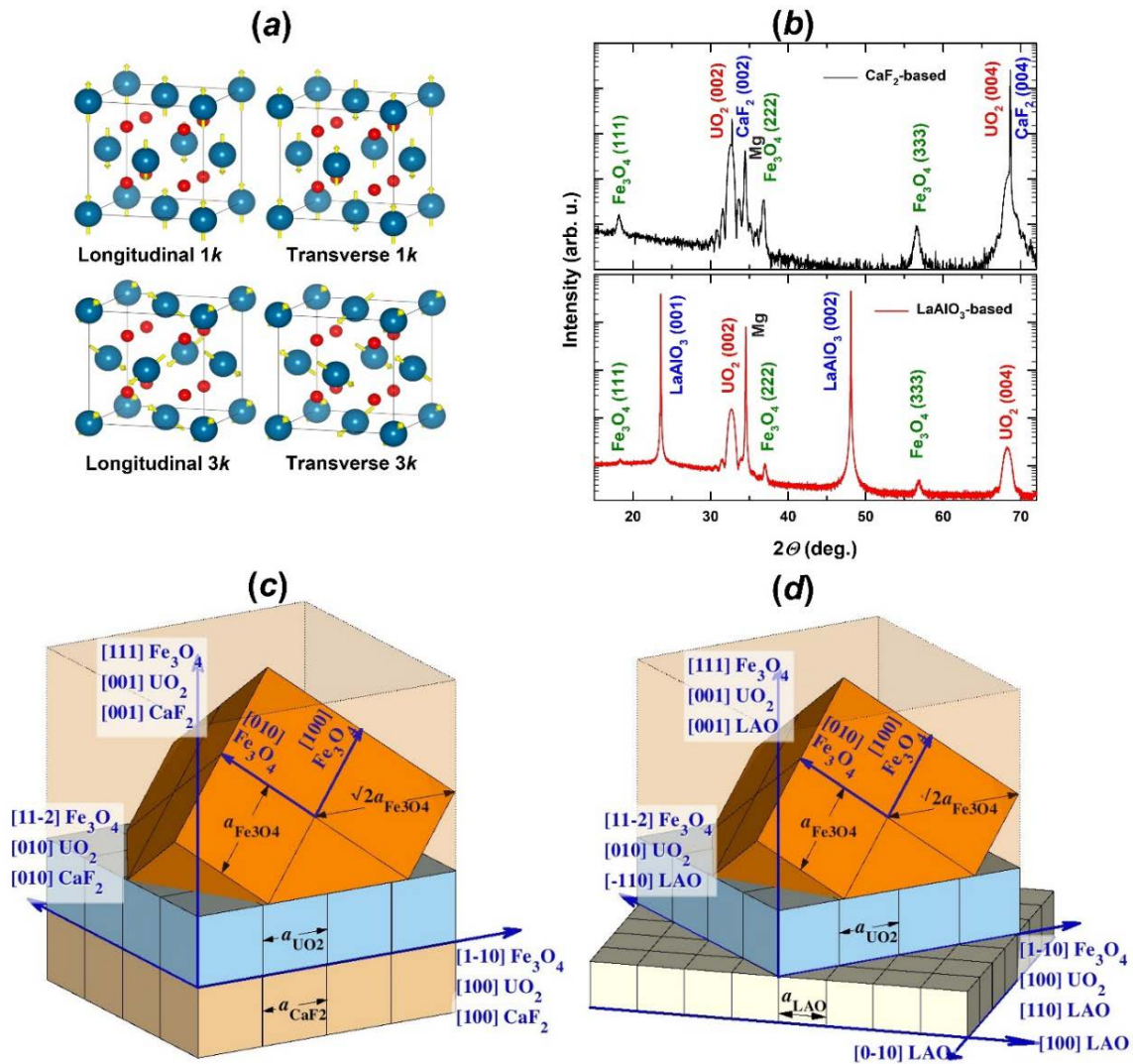

Fig. 1 (*a*) Illustration of magnetic configurations in the *fcc* crystal structure. Large symbols are magnetic atoms, small symbols are oxygen atoms. (*b*) Typical $\theta$-$2\theta$ scans for the

UO$_2$/Fe$_3$O$_4$/Mg samples on a CaF$_2$ (top) and LaAlO$_3$ (bottom) substrates. Only Bragg reflections CaF$_2$ $h00$, UO$_2$ $h00$, Fe$_3$O$_4$ $hhh$ and Mg $000l$ are visible revealing the lattice orientation for all layers. (*c*), (*d*) Epitaxial relationship found in the magnetic bilayers (111)$_{Fe3O4}$||(001)$_{UO2}$||(001)$_{LAO}$ and (111)$_{Fe3O4}$||(001)$_{UO2}$||(001)$_{CaF2}$, respectively (for details see Figs. S4, S5 and S9, Supporting information).

Magnetic hysteresis loops for both series of samples were measured after field cooling through the Néel temperature of UO$_2$ in a 10 kOe magnetic field applied in the in-plane direction of the bilayers (Fig. 2(*a*)). We determined the exchange bias effect as $H_{EB} = |(/H_{C-}| - H_{C+})|/2$, where $H_{C-}$ ($H_{C+}$) is the coercive field on the descending (ascending) branch of the hysteresis loop. The exchange bias reaches 2000 Oe in the CaF$_2$-based system. Surprisingly, we find that the LAO-based samples demonstrate a twice weaker EB as compared to the CaF$_2$-based bilayers at all thicknesses of UO$_2$ (Fig. 2(*b*)). Given that the strength of exchange coupling in the exchange-biased system is inversely proportional to the thickness of the ferro- (or ferrimagnetic in our case) magnetic counterpart of the bilayers as $H_{EB} \sim 1/\tau_F$ [27], the fact that in both systems the Fe$_3$O$_4$ layers possess similar thicknesses and have identical structural characteristics (as detailed in Figs. S4, S5, Supporting information) fails to account for the observed twofold difference in the EB magnitude.

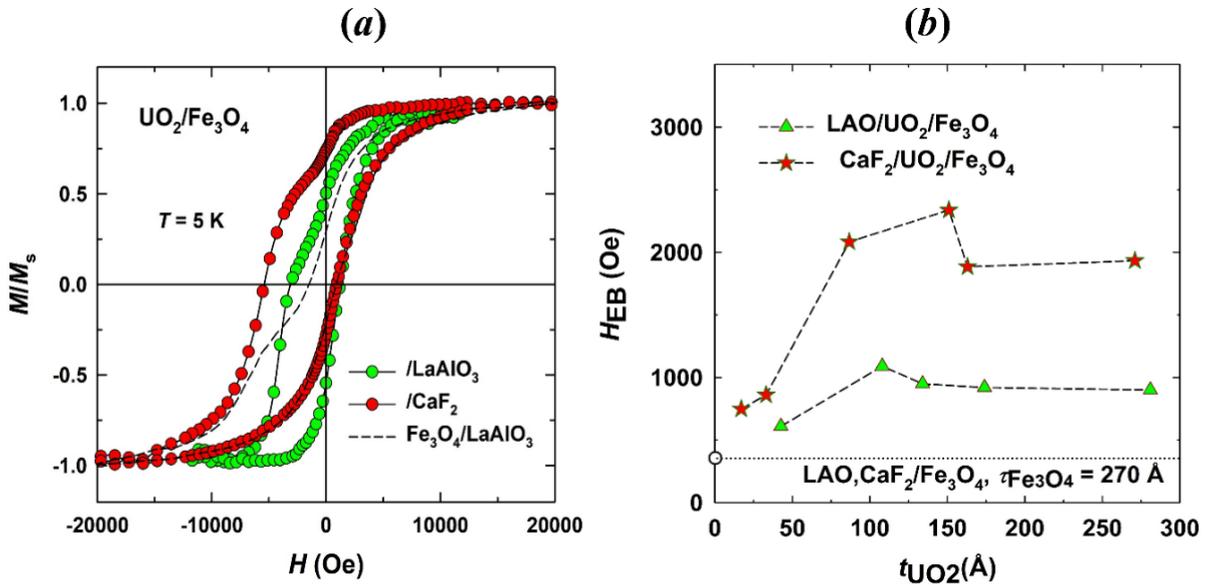

Fig. 2. (*a*) Comparison of the *H-M* magnetization loops at 5 K for the UO$_2$/Fe$_3$O$_4$ bilayers on different substrates after field cooling in the 10 kOe field. For UO$_2$ and Fe$_3$O$_4$ the thicknesses are 271 and 282 Å (CaF$_2$ substrate) and 285 and 276 Å (LAO substrate), respectively. Data for a single layer of Fe$_3$O$_4$ (270 Å) (broken line) is shown for comparison. (*b*) Dependence of exchange bias field on the thickness of the antiferromagnetic layer UO$_2$ in the UO$_2$/Fe$_3$O$_4$ bilayers. The thickness of the ferromagnetic counterpart of the bilayer, Fe$_3$O$_4$, is kept constant

in both series of the samples deposited on $CaF_2$ and $LaAlO_3$ substrates. The broken lines are a guide to the eye.

Another factor that could potentially explain the difference lies in the presence of distinct variations in the stoichiometry of the $UO_2$ layer and/or the morphology of the $UO_2/Fe_3O_4$ interface. The former was shown [26] to be responsible for ferromagnetism in the $UO_{2+x}$ thin films by incorporating point defects. Hence, the stoichiometry of the $UO_2$ layers was closely monitored after depositing each layer. In particular, we controlled the position and width of the main U-4$f$ peaks and high-energy side satellite line relative to an associated primary U-4$f_{5/2}$ peak (Fig. S1, Supporting information). This approach is independent of the absolute binding energy and is particularly useful for distinguishing the oxidation states of uranium [34]. Additionally, we ensured the absence of any foreign structures in the XPS spectra. The observed values for $UO_2$ were in line with literature [34] for both types of substrates. (As a point of comparison: in the case of hyperstoichiometric $UO_{2+x}$ thin film samples investigated in Ref. [26], the associated photoemission study documented in Ref. [35] indicated that the satellite peaks are scarcely visible.)

The quality of the interface was assessed using Rutherford Backscattering Spectrometry (RBS); the recorded well-separated spectra from U and Fe revealed no chemical intermixing between the magnetic layers within experimental resolution (Fig. S4, Supporting information). The sharpness of the interface between the $UO_2$ layer and the substrates was further validated by the presence of distinct thickness fringes observed around the $UO_2$ peaks in the XRD data (Fig. 1b and Fig. S7, Supporting information). These observations are entirely consistent with the findings from our previous transmission electron microscopy (TEM) study of the LAO-based $UO_2/Fe_3O_4$ bilayers [32].

To provide a more in-depth characterization of the $UO_2/Fe_3O_4$ samples' crystal structure, we recorded Reciprocal Space Maps in the vicinity of characteristic symmetric and asymmetric peaks of $UO_2$ on both types of substrates (see Fig. 3). The simulated lattice parameters of the $UO_2$ layer are given in Fig. 4. Our results reveal a negligible lateral strain in the $UO_2$ homogeneous epitaxial layer when deposited on $CaF_2$. The $UO_2$ layer on the LAO substrate is in-plane compressively strained. The substrate-induced tetragonality or strain of $UO_2$ (with its measure depicted in the upper part of Fig. 4) decreases with the increasing layer thickness and approaches a certain residual value given by different thermal expansions of the substrate and the layer during the post-growth cooling. This is consistent with data obtained by Bao et al. [36] who employed resonant x-ray scattering (RXS) to study $UO_2$ thin films deposited on the LAO and CaF$_2$ substrates and observed strain-induced tetragonal distortion with a mosaicity of ~1° in the LAO-based UO$_2$ films.

For the record, in some of the CaF$_2$-based layers, we also observed a noticeable tetragonal distortion, most likely induced during the cooling process after film deposition. The exchange bias in these samples immediately decreased to values similar to those observed in the LAO-based system (Fig. S10, Supporting information). On the other hand, in one of the CaF$_2$-based samples, we noticed a significantly smaller lattice parameter of UO$_2$ with $\tau_{AF} = 85$ Å (Fig. 4), probably due to a slight oxygen deficiency in the sample. However, the UO$_2$ layer in this sample remained cubic, and the resulting exchange bias was found to be of similar magnitude, around 2000 Oe, as in the stoichiometric samples. This aligns with the general understanding of the robustness of antiferromagnetism in bulk UO$_2$: bulk off-stoichiometric UO$_{2\pm x}$ remain antiferromagnetic for small deviations of $x < 0.07$ [37,38].

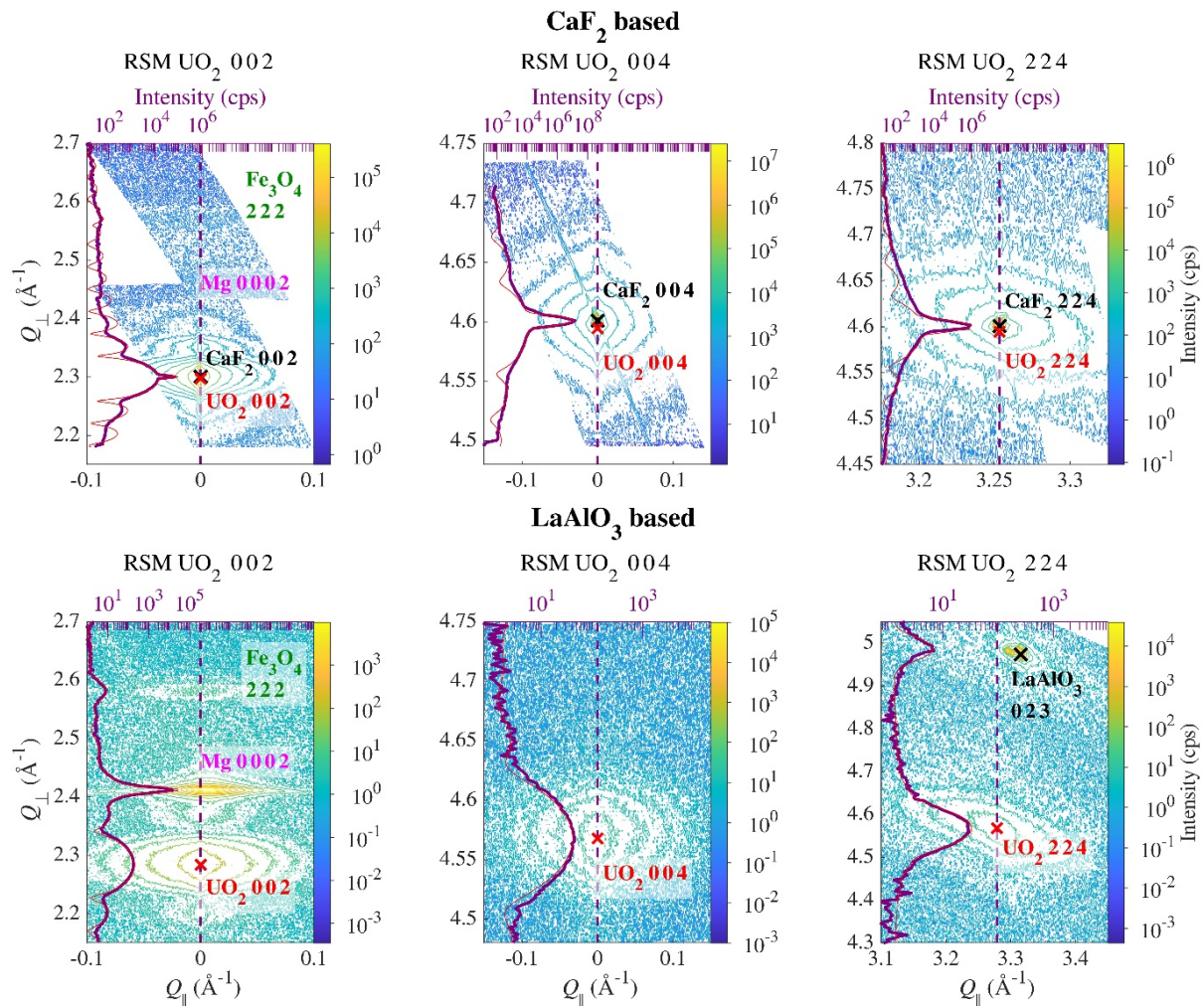

Fig. 4. Top: Reciprocal space maps in the vicinity of the UO$_2$ 002, 004 and 224 Bragg reflections. The experimental diffraction curve (magenta solid line) was extracted as a vertical cut passing through the UO$_2$ Bragg reflection and then fitted by a numerical simulation based on the kinematic theory of diffraction for determination of the out-of-plane lattice parameter

and the thicknesses for the CaF$_2$-based UO$_2$/Fe$_3$O$_4$ bilayers capped with Mg. Bottom: The same for the LaAlO$_3$-based sample.

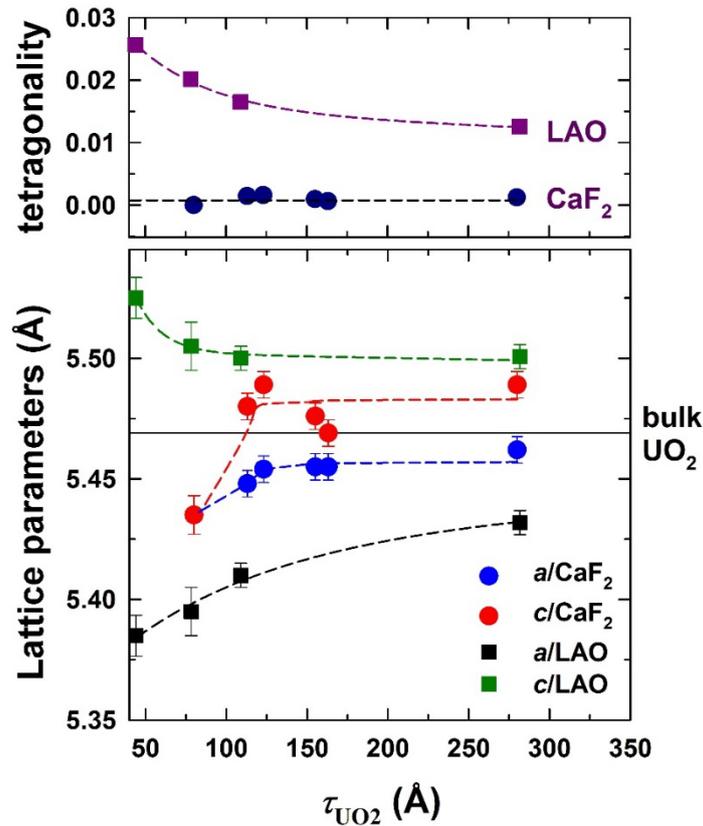

Fig. 5. UO$_2$-thickness evolution of the out-of-plane, $c$, and the in-plane, $a$, lattice parameters determined from the RSM measurements. The solid line is the lattice parameter of bulk cubic UO$_2$. The upper panel shows the strain-induced degree of tetragonality of the films determined as $(c-a)/((c+a)/2)$.

Below, we show by first-principles calculations that even a slight tetragonal distortion of the initial *fcc* structure can induce a single-ion anisotropy in strained UO$_2$. Consequently, the altered arrangement of moments within the film's bulk and at the interface could account for the discrepancy in observed exchange bias magnitudes between the two systems.

2. *Magnetic structure of the tetragonally distorted UO$_2$*

The principal possibility for a magnetic structure switching in tetragonally distorted UO$_2$ stems from the frustration between various antiferromagnetic (AFM) orders in its bulk magnetic structure, as explained in the introduction. The resolution of this frustration in cubic UO$_2$ (Fig. 1(*a*)) cannot be ascribed to the single-ion magnetic anisotropy (SIA) since the ground state of U$^{4+}$ ion in a cubic environment is a completely isotropic triplet [15]. Recent first-principles calculations [24,25] employing a correlated treatment of the 5*f*-electron shell have shown that a small energy difference between $1k$, $2k$ and $3k$ ordering is due to high-rank quadrupole-quadrupole interactions. In particular, Ref. 24 employs the DFT+HI method together with the FT-HI approach to derive the full super-exchange Hamiltonian of $UO_2$. This work showed that the experimental $3k$ is stabilized with respect to the competing structures by quadrupole-quadrupole super-exchange interactions (SEI).

In the case of tetragonally distorted $UO_2$, the SEI Hamiltonian needs to be supplemented by a SIA term:

$$H = H_{SEI} + H_{SIA} = H_{DD} + H_{QQ} + H_{SIA}, \qquad (1)$$

where $H_{SEI}$ and $H_{SIA}$ are the SEI and SIA contributions, respectively. The former is split between the dipole-dipole (DD) and quadrupole-quadrupole (QQ) terms acting within the ground-state (GS) $\Gamma_5$ triplet of $U^{4+}$ ion as defined in Ref. 23. The tetragonal crystal field (CF) splits the GS triplet inducing the SIA. We employ *ab initio* $H_{SEI}$ and $H_{SIA}$, please see Method Section for details of our theoretical approach.

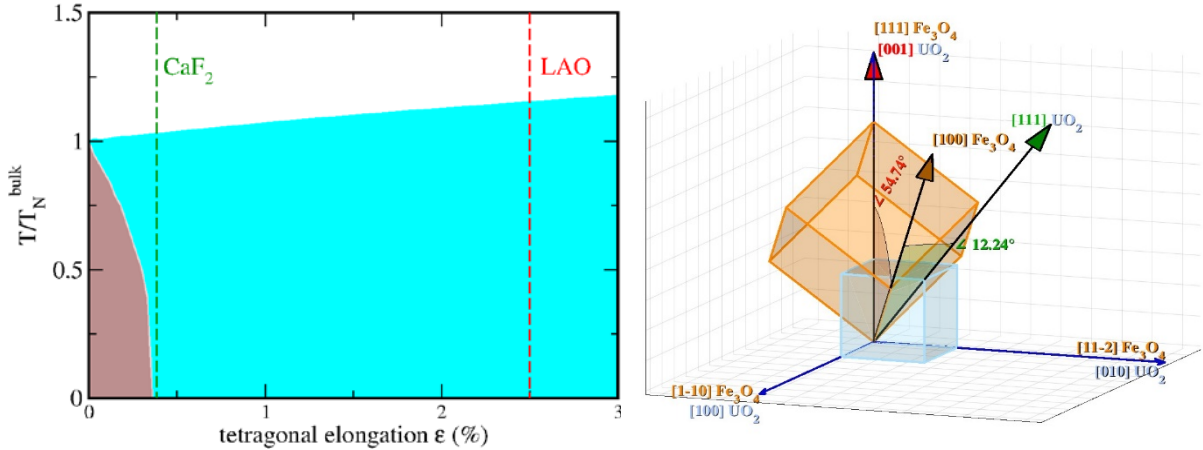

Fig. 5. (Left) Calculated magnetic phase diagram of $UO_2$. The cyan, brown and white areas display the regions of stability for the transverse $1k$, transverse $3k$ and paramagnetic phases, respectively. The temperature is related to the theoretical mean-field bulk $T_N = 56$ K. (Right) Schematic representation of various directions of magnetic moments of $UO_2$ and $Fe_3O_4$ and the corresponding angles between the moments.

For the distortion level of the LAO-based layer ($\varepsilon = c/a - 1 = 2.5$ %), we obtained the value of 210 meV for the overall crystal field (CF) splitting of the $U^{4+}$ atomic multiplet $^3H_4$. This value is very close to our previously obtained value of 207 meV for the cubic case [24] and is in good agreement with experimental estimates [15]. The $\Gamma_5$ triplet ground state is split by the tetragonal elongation into a doublet ground state and an excited singlet. This splitting is linear with $\varepsilon$ and reaches 2.78 meV for the LAO-based layer. Introducing the pseudo-angular momentum $J = 1$ to label the $\Gamma_5$ eigenstates as specified in Ref. 24, one may write the SIA term at a given U site as $H_{SIA} = D\, J_z^2$. The anisotropy constant $D = C\varepsilon$ is linear vs. the distortion, with the pre-factor $C = -113$ meV extracted from DFT+HI calculations.

We then solved the calculated *ab initio* magnetic Hamiltonian (1) as a function of the tetragonal elongation $\varepsilon$ and temperature within the mean-field (MF) approximation [39]. The resulting phase diagram is depicted in Fig. 5. Cubic $UO_2$ is predicted to order into the transverse 3$k$ AFM structure shown in Fig. 1(*a*) at the theoretical $T_N^{bulk} = 56$ K, the overestimation as compared to the experimental value of 30 K is due to the MF approximation [24]. This structure remains stable at small tetragonal distortions, however, the moment amplitude along the out-of-plane *z* direction increases with $\varepsilon$ due to the SIA effect. The ordered moment at U site $\boldsymbol{R}$ is given $M^{\boldsymbol{R}} = \left[M_{\|}\exp(-i\boldsymbol{k}_z\boldsymbol{R}), M_{\|}\exp(-i\boldsymbol{k}_x\boldsymbol{R}), M_{\perp}\exp(-i\boldsymbol{k}_y\boldsymbol{R})\right]$, where $M_{\|}$ and $M_{\perp}$ are the amplitude of the in-plane and out-of-plane components of the moment, respectively, $M_{\|} < M_{\perp}$, and $\boldsymbol{k}_{x,y,z}$ are propagation vectors in the corresponding directions. At the transition point between the two structures the in-plane amplitude $M_{\|}$ becomes 0, thus the structure collapses into the transverse 1$k$ one (also depicted in Fig. 1(*a*)) with the out-of-plane moment direction. We note that the actual distortion level for $CaF_2$-based films is very close to the 1$k$→3$k$ transition boundary. Provided the sensitivity of this boundary to the calculational details and experimental conditions, the $CaF_2$-based layers can still be considered to be in the 3$k$ domain, while the LAO layers are firmly situated within the 1$k$ region.

*3. Discussion*

Based on our first principles calculations, one may intuitively expect that different mutual arrangements of magnetic moments of cubic and tetragonally distorted $UO_2$ and $Fe_3O_4$ would give different EB values since the interface exchange interaction creating the EB pinning would also essentially change. The situation is illustrated in a simple cartoon in Fig. 5(left). If the magnetic structure of $UO_2$ remains 3$k$ in the $CaF_2$-based thin films, the magnetic moments of $UO_2$ lying in one of the three equivalent <111> directions couple with the magnetic moment of $Fe_3O_4$ lying along the <100> directions. The angle between the moments directions is 12.24°. In this case, we observe the maximum EB effect in the $CaF_2$-based samples. If switching to another AFM state occurs upon straining the lattice of $UO_2$, then the coupling strength and hence the magnitude of EB essentially change. Our calculations predict stabilization of a **1$k$**-AFM structure (Fig. 5) in the distorted crystal structure, hence the orientation of magnetization of the AFM sublattices would then lie in the [001] direction. In this case, coupling with the moment of $Fe_3O_4$ occurs at the angle 54.74°, resulting in the drastic reduction of EB for the LAO-based samples.

In general, it is assumed that the bulk spin configuration of a material is maintained even at the interface. However, real exchange-biased systems are much more complex than simple models. The orientation of interfacial spins can be influenced by various factors, such as structural properties like crystallinity and morphology, which often manifest themselves very differently for various material combinations. As a result, the interfacial spin orientations in such systems can be more intricate and difficult to predict. For the LAO-based samples, in addition to reduced exchange interaction due to differently oriented moments of the bulk $UO_2$ layer, larger mosaicity [36] of the strained $UO_2$ layer may also contribute to the decreased $H_{EB}$.

It is worth noting that tetragonal deformation appearing as the result of compression of $UO_2$ deposited on a LAO substrate in Ref. 36 was shown to promote a formation of a magnetically "dead" layer close to the substrate. The exact nature of the "dead" layer was not determined in the RXS experiment, but the evidence was such it did not have the antiferromagnetic structure of bulk $UO_2$. As it follows from our combined XRD, XRR (Supporting information) and RBS studies, the interface between the substrate and the $UO_2$ layer for both types of substrates is well-defined for all $UO_2$ layer thicknesses. In our $UO_2/Fe_3O_4$ bilayers, we detect exchange bias in both systems even at the smallest $UO_2$ thicknesses. For both types of samples, the EB magnitude and hence, the interfacial interaction and exchange coupling become less prominent at $UO_2$ thicknesses below 70 Å, which is in line with general behavior for various exchange-biased systems [40]. The peculiar feature of our samples is that EB never falls to zero completely. It is related to the presence of non-zero exchange bias in the magnetite layer [32]. From studying the control samples of single $Fe_3O_4$ layers shown in Fig. 2(*a*) with identical structure and similar thicknesses to that of bilayers, we find that a certain part of the effect (~15% at 5 K in the $CaF_2$-based samples) is naturally coming from magnetite alone. We emphasize that the saturation magnetization in the single $Fe_3O_4$ layers and in the bilayers is essentially the same. (Low-temperature magnetic moment of about 440 emu/cc is found for both types of films, which is slightly lower than the saturation magnetization value of 480 emu/cc of bulk magnetite [41].) Importantly, the magnitude of exchange bias in our $UO_2/Fe_3O_4$ bilayers on both types of substrates always remains significantly larger than that of a single magnetite layer of similar thickness.

3. Conclusion

The exchange bias magnitude in the stoichiometric, homogeneous $UO_2/Fe_3O_4$ bilayers depends on the type of substrate used. In case of small tetragonal distortion in the LAO-based $UO_2$ layers, the EB in bilayers is significantly reduced as compared to the $CaF_2$-based samples. By first principles calculations we show that the reason dwells in the change of antiferromagnetic structure as a result of distortion. $UO_2$ films with a cubic $CaF_2$ structure are very close to the border of 3***k***-1***k*** transition and strain imposed by the substrate pushes $UO_2$ toward the 1***k*** antiferromagnetic structure. Our study reveals a prospect of switching frustrated magnets for potential applications. For instance, by pairing $UO_2$ with a soft ferromagnet (permalloy, Py) we were able to induce perpendicular magnetic anisotropy in the Py layer through the canted moments of a $CaF_2$-based $UO_2$ [42]. The effect might be greater with a LAO-based $UO_2$, where the direction of $UO_2$ and Py moments would coincide. The inherent ability to manipulate the magnetic alignment in frustrated magnets opens avenues for tailoring device functionalities to specific requirements.

## 4. Method Section

**Experiment**

The $UO_2$/$Fe_3O_4$ bilayers were synthesized by reactive (-gas) *dc* sputtering realized in a home-built setup, using a miniature U (natural uranium, 99.9 wt.% purity) and Fe (99.99% purity) targets and an electron emitting thoriated tungsten filament stabilizing the plasma [43]. Different epitaxial strain states in the $UO_2$ layer were realized by employing different substrates. The thickness of the $UO_2$ layer varied across the samples series, with the minimal thickness of 17 Å on the $CaF_2$ substrate and with the maximum of 340 Å on $SiO_2$. In contrast, the thickness of the $Fe_3O_4$ layer in the bilayers was maintained at approximately 270 ± 20 Å. Prior to the growth, the commercially available substrates $CaF_2$ (100), $LaAlO_3$ (001) and fused silica $SiO_2$ (MTI Corp.) (the results for the latter are shown in Figs. S9 and S10, Supporting information) were annealed in dynamic vacuum at temperatures used for $UO_2$ deposition as shown below. The $UO_2$ layer was always deposited onto the substrates at elevated temperature of 350 ºC for $CaF_2$ and $SiO_2$ substrates and at 500 ºC for the LAO substrate, using a partial oxygen pressure of $1.2 \times 10^{-6}$ mbar (Ar pressure of $6 \times 10^{-3}$ mbar). We specifically restricted the deposition temperature for the $UO_2$ layers to 500 ºC for LAO substrates to avoid passing through the ferroelastic transition that occurs at 560 ºC in this substrate. As for the $CaF_2$ substrates, we have chosen even lower deposition temperatures to avoid cracking of the substrate when cooling from growth temperature to ambient conditions.

The layer of $Fe_3O_4$ was deposited on the top of $UO_2$ at room temperature to avoid interdiffusion and at $6\text{-}7 \times 10^{-7}$ mbar oxygen pressure (Ar pressure of $1\text{-}3 \times 10^{-2}$ mbar). The stoichiometry of each deposited layer was controlled *in situ* by X-ray Photoelectron Spectroscopy (XPS) using a MgKα (photon energy 1253.6eV) source. After that, a magnesium cap was deposited for protection on top of each sample. For the sake of comparison of properties, we also prepared 270-Å thick single $Fe_3O_4$ layers on the same substrates and capped them with Mg, too.

The films' structure was characterized by several methods. We used Rutherford Backscattering Spectrometry (RBS) in the channeling mode (RBS-C) using a beam of 2-MeV $He^+$ ions to check the single-crystalline layer structure and to determine the sharpness of the interface between the magnetic layers. RBS-C measurements were carried out on a Tandetron MC 4130 accelerator at the Nuclear Physics Institute of the Czech Academy of Sciences in Řež near Prague. For RBS-C analysis, the crystal target was mounted onto a two-axis goniometer with an angular resolution of 0.01°. A surface-barrier Si detector with a circular aperture of 5.4 mm in diameter was placed at a distance of 70 mm from the sample holder at the scattering angle of 170°. The energy resolution of the detector was about 12 keV. The accessible information depth provided by 2-MeV $He^+$ ion beam is about ∼1.5 μm in this geometry. The spectra were evaluated with the software SIMNRA 6.06 [44].

The x-ray reflectometry (XRR) and x-ray diffraction (XRD) measurements were performed with Rigaku SmartLab diffractometer equipped with a 9kW rotating anode. The primary beam was monochromatized and parallelized by a parabolic x-ray mirror and a 2-bounce Ge(220) monochromator. Reflectivity curves (Fig. S6, Supporting information) and XRD $\theta$-$2\theta$ scans were measured with two receiving slits in front of the detector and the XRD reciprocal space map (RSM) measurements were performed with a HyPix-3000 in the 1D mode (medium resolution). From the XRR data we determined the thickness of individual layers and roughness of the corresponding interfaces. Lattice parameters and a relaxation state of the $UO_2$ layers was found using RSM measurements allowing us to obtain an out-of-plane lattice parameter and the thickness of the single crystal layers, with precise values determined by simulation based on the kinematic theory of diffraction.

The magnetization studies were conducted using a vibrating sample magnetometer on the Quantum Design PPMS9 platform (Quantum Design). The magnetic field was applied along the sample surface.

**Theory**

We employ the same SEI as calculated in Ref. 24 for cubic $UO_2$, thus neglecting the effect of tetragonal distortions on their magnitude, which is likely unimportant provided the rather small distortion magnitude. In order to evaluate the SIA term, we calculate the splitting of the GS triplet in $UO_2$ versus the magnitude of tetragonal distortions by means of the self-consistent DFT+Hubbard-I (HI) approximation [28-30] using the same calculational parameters as in Ref. 24. Namely, we employ $U = 4.5$ eV and Hund's rule coupling $J_H = 0.6$ eV to specify a rotationally invariant on-site Coulomb repulsion on the U 5$f$ shell, the fully-localized limit double-counting term for the nominal atomic occupancy of 2 of the U$^{4+}$ 5$f$ shell. 3000 $\boldsymbol{k}$-points were employed in the Brillouin zone integration and the LAPW basis cutoff [29] $R_{MT} \cdot K_{MAX} = 7$. We adjusted the out-of-plane lattice parameter $c$ so as to keep the cell volume fixed at its bulk value.

**Supporting information**

Supporting information is available from the Wiley Online Library or from the author.

**Acknowledgements**

We would like to thank Dr. G. H. Lander for valuable discussions during the course of this work. We acknowledge the support of Czech Science Foundation under the grant no. 22-19416S. The samples were prepared in the framework of the EARL project of the European Commission Joint Research Centre, ITU Karlsruhe. RBS measurements were carried out at the CANAM infrastructure of the NPI CAS Rez supported through MEYS project No. LM2015056. Physical properties measurements were performed in the Materials Growth and Measurement Laboratory (http://mgml.eu/) supported within the program of Czech Research Infrastructures (project no. LM2018096). L.V.P. acknowledges support from European Research Council Grant ERC-319286-"QMAC" and the computer team at CPHT.

**Conflict of interest**

The authors declare no conflict of interest.

**Data Availability Statement**

The data that support the findings of this study are available from the corresponding author upon reasonable request.